\documentclass[conference]{IEEEtran}
\IEEEoverridecommandlockouts

\usepackage{amssymb}
\usepackage[cmex10]{amsmath}
\usepackage{stfloats}
\usepackage{graphicx}
\usepackage{subfigure}
\usepackage{tabularx}
\usepackage{epsfig,epsf,color,balance,cite}
\usepackage{verbatim}
\usepackage{url}
\usepackage{bm}

\newtheorem{theorem}{\bf Theorem}

\usepackage{algorithm}
\usepackage{algorithmic}

\usepackage{color}
\definecolor{myc1}{rgb}{0,0,0}

\begin{document}

\title{{Resource Allocation for UAV Assisted Wireless Networks with QoS Constraints}  }

\author{
\IEEEauthorblockN{Weihang Ding\IEEEauthorrefmark{1},
Zhaohui Yang\IEEEauthorrefmark{1},
Mingzhe Chen\IEEEauthorrefmark{2},
Jiancao Hou\IEEEauthorrefmark{1}
                  and Mohammad Shikh-Bahaei\IEEEauthorrefmark{1}
                  }
\IEEEauthorblockA{\IEEEauthorrefmark{1}Centre for Telecommunications Research, Department of Engineering, King's College London, London WC2R 2LS, UK.}
\IEEEauthorblockA{\IEEEauthorrefmark{2}Electrical Engineering Department of Princeton University, USA, \\
The Chinese University of Hong Kong, Shenzhen, China.}

\vspace{-2em}
}

\maketitle

\begin{abstract}
For crowded and hotspot area, unmanned aerial vehicles (UAVs) are usually deployed to increase the coverage rate. In the considered model, there are three types of services for UAV assisted communication: control message, non-realtime communication, and real-time communication, which can cover most of the actual demands of users in a UAV assisted communication system. A bandwidth allocation problem is considered to minimize the total energy consumption of this system while satisfying the requirements. Two techniques are introduced to enhance the performance of the system. The first method is to categorize the ground users into multiple user groups and offer each group a unique RF channel with different bandwidth. The second method is to deploy more than one UAVs in the system. Bandwidth optimization in each scheme is proved to be a convex problem. Simulation results show the superiority of the proposed schemes in terms of energy consumption.
\end{abstract}

\IEEEpeerreviewmaketitle

\vspace{-0.5em}
\section{Introduction}\vspace{-.5em}

With the prevalent using of mobile communication devices, the conventional wireless networks are greatly challenged nowadays. The normal ground base stations may be inadequate if there are disasters or unexpected big crowds in the region. Besides, some of the regions can hardly be served by ground base stations because of hardness of facilities construction or huge obstacles such as skyscrapers and mountains \cite{Zhao}. In these cases, unmanned aerial vehicles (UAVs) assisted wireless communication platform is a rather promising technology \cite{Gomez}. 

The major advantage of UAV-assisted wireless communication is that UAVs have high altitudes, which result in the high qualities of the channels. The probability that the channels between the ground users and the UAV are line of sight (LoS) increases as the altitude of the UAV increases. The channel quality also depends on ambient environmental parameters and the horizontal distance between the user and the UAV \cite{P1,P2,P3}. Significant efforts have been applied for this field based on the channel model. The optimal three-dimensional (3D) position distribution of the UAVs is studied in \cite{Al-Hourani,Lyu,Chen}. Other aspects of UAV-assisted wireless communication network are also studied previously such as coverage analysis \cite{Chetlur}, \cite{Coverage}, cell partitions of the UAVs \cite{cell}, and spectrum sharing \cite{chen1}. The second type of UAV-assisted wireless communication employs mobile UAVs. These UAVs can serve as not only base stations, but also relays \cite{relay}. As mobility is one of the most important property of the second type of UAVs, their trajectories are needed to be determined. The UAV's trajectory can be periodic or non-periodic based on the actual requirements to optimize the overall performance of the system \cite{trajectory}, \cite{chen2}. Moreover, there are many previous works focusing on other parameters of UAVs to optimize the performance of the communication system in slightly different scenarios. These parameters include power \cite{power}, beamwidth \cite{beamwidth}, bandwidth assignment \cite{bandwidtha}, etc. 

In order to optimize the performance of a  UAV-assisted wireless communication system, the overall energy consumption is one of the most critical issue owing to the limited budget of the battery capacity. Apart from energy consumed by wireless communication, certain amount of propulsion energy is used by the UAV to support their movements. As a result, the whole system ought to be designed differently from the ground system. The power consumption model for helicopters is defined in the well known textbook \cite{Bramwell}. In \cite{tradeoff}, the authors studied the trade-off between the propulsion energy and communication related energy consumed by the system to find the optimal transmit power of the fixed ground users and the trajectory of the fixed-wing UAV. As the UAV gets closer to the ground user groups, the energy cost to transmit messages can be significantly reduced while additional propulsion energy is needed to enable the UAV's movement.The main aim of the UAV discussed in \cite{tradeoff} is collecting data from ground terminals. Quality of Service (QoS) is a significant issue to be considered in UAV-assisted communication systems. The author in \cite{QoS1} and \cite{QoS2} focused on the 3-D placement of UAVs to achieve the required QoS with the help of multiple algorithms.

However, most of the previous works merely focused on the energy used for transmitting signals and ignore the propulsion energy of the UAVs. The total energy consumed by the same system when served by different numbers of UAVs was rarely compared by previous works. Moreover, the above works ignore the joint consideration of energy consumption for multiple UAVs and the bandwidth allocation for multiple users. An iterative algorithm is proposed in \cite{yang} which results in higher energy efficiency and lower delay compared to conventional federated learning (FL) methods.

The objective of this paper is to study the bandwidth assignment and the UAV deployment for this UAV assisted wireless communication network. We aim to minimize the overall energy consumed by this whole system while the available bandwidth is assigned to different radio frequency (RF) channels to satisfy different types of QoS.

The main contributions of this paper are listed as follows:

\begin{itemize}
	\item The optimal assignments of bandwidth to different users and services under energy limitations are obtained.
	\item The users are categorized based on their channel qualities for the purpose of saving energy.
	\item The optimal UAV distribution is studied through the comparison between the system served different amounts of UAVs.
\end{itemize}

The rest of this paper is organized as follows. The system model is described in Section II. The problem formulation is described in Section III. Section IV and V are the numerical results and conclusion of this paper.

\vspace{-.5em}
\section{System Model}
\vspace{-.25em}

\subsection{RF Channel Model}
Consider an UAV-assisted wireless communication system with $n$ ground users uniformly distributed in a square area with length $R$. The horizontal positions of the ground uses are denoted by the set $\mathcal{X} = \{\boldsymbol{x_i}\}_{i=1}^n = \{(x_i,y_i)\}_{i=1}^n$. There are $k$ UAVs deployed in this system. The 3-D positions of UAVs can be denoted as set $\mathcal{Y} = \{\boldsymbol{y_j}\}_{j=1}^k = \{(x_j^0,y_j^0,h_j^0)\}_{j=1}^k$.
Let $\mathcal I_j$ denote the set of users served by UAV $j$. To avoid collision among the UAVs, the distance between any two UAVs must not exceed $R_{max}$, i.e., $\| \boldsymbol{y_{m}}-\boldsymbol{y_{n}}\| \ge R_{max}$, $\forall m,n \in\{1, \cdots, k\}$.

The overall channel gain (expressed in dB) between user $i$ and UAV $j$ can be modeled as:

\begin{equation}
g_{i,j}=-\underbrace{\left(20\log{d}+20\log{f_c}+20\log{\frac{4\pi}{c}}\right)}_\text{Free space pass loss}-\eta
\end{equation}
where $d=\sqrt{{h_j^0}^2+(x_{i}-x^0_j)^2+(y_{i}-y^0_j)^2}$, $f_c$ is the carrier frequency of the RF channel, and $\eta$ here refers to the excessive pass loss determined by whether the channel is LoS and the ambient environment.

 The probability of LoS, which is closely related to the ambient environment and the density and distribution of large obstacles including high buildings and mountains, can be defined as:
\begin{equation}
P_{LoS,i,j}=\frac{1}{1+a \exp{
\left(-b\left(\frac{180}{\pi}\tan^{-1}\left(\frac{h_j^0}{d_{H,i,j}}\right)-a \right) \right)} 
}
\end{equation}
where $d_{H,i,j}=\sqrt{(x_{i}-x_j^0)^2+(y_{i}-y_j^0)^2}$ denotes the horizontal distance between user $i$ and UAV $j$, $a$ and $b$ are environmental factors.\cite{Mozaffari} As the channel is either Los or non LoS (NLoS), the probability of NLoS can be written as
\begin{equation}
P_{NLoS,i,j}=1-P_{LoS,i,j}
\end{equation}

Accordingly, we can rewrite $g_{i,j}$ as:

\begin{equation}
g_{i,j}=\left\{
\begin{aligned}
&\-(20\log{d}+20\log{f_c}+20\log{\frac{4\pi}{c}})-\eta_{LoS} \text{,}\\
&\:\:\:\:\:\:\:\:\:\:\:\:\:\:\:\:\:\:\:\:\:\:\:\:\:\:\:\:\:\:\:\:\:\:\:\:\:\:\:\:\text{if the channel is LoS}\\
&\-(20\log{d}+20\log{f_c}+20\log{\frac{4\pi}{c}})-\eta_{NLoS} \text{,}\\
&\:\:\:\:\:\:\:\:\:\:\:\:\:\:\:\:\:\:\:\:\:\:\:\:\:\:\:\:\:\:\:\:\:\:\:\:\:\:\:\:\text{if the channel is NLoS}
\end{aligned}
\right.
\end{equation}
where $\eta_{LoS}$ and $\eta_{NLoS}$ are the excessive pass loss factor of LoS and NLoS channels respectively \cite{Al-Hourani}.

The average expected channel gain of a user located at $(x_i,y_i)$ can be given by:
\begin{equation}
\begin{aligned}
\mathbb{E}[g_{i,j}]&=P_{LoS,i,j}g_{LoS,i,j}+P_{NLoS,i,j}g_{NLoS,i,j}=\bar{g}_{i,j}
\end{aligned}
\end{equation}

The transmit power of messages depends on the channel quality, available bandwidth, and required channel capacity. The average power and energy costed by the wireless channel between user $i$ and UAV $j$ in this system can be written as:

\begin{equation}
P_{ij}=\left(2^{\frac{C}{b_\xi}}-1\right)\frac{N_0b_\xi }{\bar{g}_{i,j}}
\end{equation}

\begin{equation}
E_{ij}=P_{ij} T_{tran}=\left(2^{\frac{C}{b_\xi}}-1\right)\frac{N_0b_\xi T_{tran}}{\bar{g}_{i,j}}
\end{equation}
where $C$ is the data rate between the UAV base station $j$ and the ground user $i$, $b_\xi$ denotes the bandwidth allocated to the $\xi$th RF channel, $N_0$ is the spectral density of Additive white Gaussian noise (AWGN), and $T_{tran}$ is the overall transmission time.

\subsection{Propulsion Energy Model}

According to textbooks \cite{Bramwell} and \cite{Filippone}, the UAV's torque coefficient $q_c$ while hovering can be expressed as

\begin{equation}
q_c=\frac{\delta}{8}\left(1+\frac{3V^2}{\Omega^2R_u^2}\right)+\left(1+k\right)\frac{T_u\lambda_i}{\rho sA_u\Omega^2R_u^2}+\frac{1}{2}d_0\frac{V^3}{\Omega^3R_u^3}
\end{equation}
where $\delta$ is profile drag coefficient, $V$ is the UAV's speed, $k$ is the incremental correction factor to induced power, $\rho$ is the air density, $R_u$ is the radius of the UAV's rotor, $\Omega$ is the angular velocity of the rotors, and $T_u$ is the rotor thrust of this UAV. $A_u$ denotes the disc area covered by the UAV's rotor which can be calculated from equation $A_u=\pi R_u^2$. The rotor solidity $s$ is defined as the ratio of the overall rotor area to the rotor disc area. In other words, it denotes how much of the rotor disc area is covered by the rotor blades at a time. $s$ can be expressed as $s=\frac{bc}{\pi R_u}$ where $b_u$ and $c_u$ are the number of rotor blades and the chord length of the rotor blades. 
$\lambda_i$ denotes the mean induced velocity normalized by the UAV's tip speed. It is described more detailed later in this paper. 
$d_0$ is the UAV's fuselage drag ratio. While a UAV is moving at high speed, air resistance can be a significant resource of energy cost. A UAV's fuselage drag ratio $d_0$ in this case is defined as $d_0=\frac{S_{FP}}{sA_u}$, where $S_{FP}$ is the UAV's property defined as the fuselage equivalent flat plate area.

The relationship between a UAV's required power and torque coefficient is 

\begin{equation}
\begin{aligned}
P_{0}&=q_c\rho sA_u\Omega^3R_u^3\\
&=\frac{\delta}{8}\rho sA_u\Omega^3R_u^3\left(1+\frac{3V^2}{\Omega^2R_u^2}\right)+(1+k)T_uv_{i0}\\
&+\frac{1}{2}d_0\rho sA_uV^3
\end{aligned}
\end{equation}
where $v_{i0}$ is the mean induced velocity and it can be obtained by equation $v_{i0}=\lambda_i\Omega R_u$. Furthermore, in the case where the UAV operates with a speed $V$ and a rotor thrust $T_u$, the mean induced velocity of this UAV $v_{i0}$ can be calculated as follows

\begin{equation}
v_{i0}=\sqrt{\sqrt{\frac{T_u^2}{4\rho^2A_u^2}+\frac{V^4}{4}}-\frac{V^2}{2}}
\end{equation}

In our system model, we assume that the UAVs are static and, hence, their velocity $V=0$, and its rotor thrust $T_u$ equals to the UAV's weight, which is denoted by $W$. Besides, in our system, the energy used by the UAVs to reach their designated position is ignored. Therefore, the overall propulsion power of this UAV assisted wireless communication system can be expressed as

\begin{equation}
P_{UAV}=\frac{\delta}{8}\rho sA_u\Omega^3R_u^3+(1+k)\frac{W^{3/2}}{\sqrt{2\rho A_u}}
\end{equation}

\subsection{Service Types}

In order to improve the utilization of bandwidth and energy and avoid excessive resources, all services provided by the UAV-assisted wireless communication system are categorized based on their QoS requirements. There are three types of services discussed in this paper, which can cover almost all potential user needs.
\begin{itemize}
	\item \textbf{Control message:} 
	This kind of messages are always quite short packets containing only a few numbers or commands (typically a few hundred bits) but require very low latency, which is used to transfer control messages like the instructions to UAVs, the real-time information of the ambient environment and so on. All control messages are transmitted by one single RF channel to minimize the queuing latency.
	\item \textbf{Non-real-time message:} 
	This kind of messages are used to transfer delay-tolerant files whose packets are much bigger than control messages. Therefore, limitations like latency and block probability can be ignored in this case. There are $k_n$ RF channels allocated to this kind of service. 
	\item \textbf{Real-time communication:} 
	This type of service is like Public Switched Telephone Network (PSTN). A user will occupy a channel during the connection and the call request is rejected if there is no empty channel. The number of RF channels $k_r$ used for this kind of service is based on the overall traffic load and the maximum block probability. 
\end{itemize}

\subsubsection{Control Message}

It is assumed that the time that the packets from each user are transmitted is Poisson distributed with an average transmission rate $\lambda$. The length of the packets are assumed to be exponentially distributed with an expectation of $L$. Therefore, this queuing network can be modeled as a M/M/1 queuing network with infinite buffer size. If we split the whole channel into $k$ sub-channels using Frequency Division Multiplexing (FDM) or Time Division Multiplexing (TDM), the average queuing latency is $k$ times larger than it in the system while only a whole channel is used. As our major objective is to reduce the latency of this kind of messages, and the packets are so short that the probability of collision is rather low, we will use a whole RF channel to transmit control messages. 

The data rate requirement of this channel can be calculated as

\begin{equation}
C_{min}=\frac{1}{T_{req}}+n\lambda L
\end{equation}
where $T_{req}$ is the maximum allowed latency of control messages.

CSMA/CD (listen before and while talking) protocol is used. If the user finds the channel busy, the transmission will be inhibited until the channel turns idle. However, while collisions happen, an exponential back-off algorithm is introduced to determine the re-transmission process. The major disadvantage of using CSMA/CD in this system is that the delay is more variable and harder to evaluate. But we still can find methods to find out the average delay and throughput of a CSMA/CD protocol. The utilization of CSMA/CD is $\frac{1}{1+3.44\frac{T_{prop}}{T_{tran}}}$, so the data rate requirement of this channel can be expressed as

\begin{equation}
C_{req}=\left(1+3.44\frac{T_{prop}}{T_{tran}}\right)C_{min}
\end{equation}
where $T_{tran}$ is the time needed for a packet to be transmitted through this channel and is defined as $T_{tran}=\frac{L}{C_{min}}$, and $T_{prop}$ is the time needed for the wireless signal to propagate to the correlated UAV base station. However, the signal's propagation speed in such a wireless communication system is the speed of light which is $3*10^8$ meters per second. Hence, the transmission time in this system is several hundred times longer than the propagation time, so that $C_{req}$ is very close to $C_{min}$, which means collisions in this system have got little effect. So we can ignore the collisions of packets and use the value of $C_{min}$ as $C_{req}$ in further studies.

The bandwidth that is allocated to control messages is $b_c$ and the achievable data rate in practice is $C_c$, then, the energy consumed by the channel between ground user $i$ and UAV $j$ to transmit control messages in one second can be written as

\begin{equation}
\begin{aligned}
E_{i,j,c}&=P_{i,j,c} T_{tran} \lambda=\left(2^{\frac{C_{c}}{b_c}}-1\right)\frac{N_0b_cL\lambda}{g_{i,j}C_{c}}
\end{aligned}
\end{equation}

\subsubsection{Non-real-time Message}

The multiple access technique used to transmit non-real-time messages is TDMA combined with FDMA. We split the bandwidth used for this type of service into multiple RF channels and divide each RF channel into multiple time slots. Each user who wants to transmit non-real-time messages can use only one physical channel at a time. It is assumed that $k_n$ RF channel is allocated for this type of service and the minimum allowed data rate is $R_n$.

In the conventional cellular wireless communication system, the whole bandwidth is evenly distributed into multiple RF channels. Each RF channel is selected by users randomly. 

However, in such a UAV assisted wireless communication system, the quality of the channels can vary significantly depends on whether the channel is LoS or NLoS. In this paper, we are trying to find a method to categorize the users into several groups based on the channel quality between each user and the UAV base station. The available bandwidth is distributed to each user group depending on the channel quality of users in one group. The advantage of this method compared with the traditional way can be reflected by the reduction of energy consumed in a fixed amount of time.

The users that use the $\xi$th RF channel are belong to set $\mathcal{R}_\xi$, which can be expressed as $i\in \mathcal{R}_\xi$.

The energy consumed by the transmission of non-real-time messages by the channel between ground user $i$ and UAV $j$ in each second can be written as

\begin{equation}
\begin{aligned}
E_{i,j,n}&=P_{i,j,n}T_{tran}=\left(2^{\frac{C_{\xi}}{b_\xi}}-1\right)\frac{N_0b_\xi }{g_{i,j}}
\end{aligned}
\end{equation}
where $C_\xi$ denotes the actual data rate of user $i\in \mathcal{R}_\xi$ to transmit non-real-time messages, and $b_
\xi$ is the bandwidth of the RF channel that is used by user $i\in \mathcal{R}_\xi$.

\subsubsection{Real-time Communication}

This type of service is like public switched telephone network (PSTN). Whenever a user wants to initiate a real-time communication, an idle channel will be selected and occupied during this communication. If all the channels are unavailable at this time, then the user will fail to get a connection and the communication is said to be blocked. The parameter Grade of Service (GoS) $B_l$ is defined to measure the probability that a communication is failed to set up.

It is assumed that the frequency that the communications are requested to initiated are Poisson distributed with an expected value of $Q$. It means that on average, $Q$ communications occur per unit time $t$, i.e. one hour. The average duration of the connections is $T$. Therefore, it can be seen as a M/M/m/m queuing system with no buffer size, where $m$ stands the total number of available channels assigned to this type of service. The probability of the last state in the Markov chain is the blocking probability $B_l$ in this system. 

The minimum required channels in the system can be obtained by the following Equation (\ref{Erlang}) or Erlang-B tabular. Erlang is the unit of traffic load $A$ and can be calculated by $A=\frac{QT}{t}$. The Erlang-B can be given by:

\begin{equation}
\label{Erlang}
B_l=\frac{A^N/N!}{A^0/0!+A^1/1!+A^2/2!+\cdots+A^N/N!}
\end{equation}
where $N$ denotes the number of physical channels used for this type of service.

Each RF channel can be divided into ten subframes with burst data rate $R_b$ according to Long Term Evolution (LTE) frame structure, so the number of RF channels required for real-time communication can be calculated by $k_r=N/10$.

About 100 ms of latency is tolerable for real-time service. Therefore, latency limitation can be ignored for real-time communication. The energy consumed by the channel between ground user $i$ and UAV $j$ for real-time communication in each second can be expressed as

\begin{equation}
\begin{aligned}
E_{i,j,r}&=P_{i,j,r} T_{tran}=\left(2^{\frac{C_{r}}{b_{r}}}-1\right)\frac{N_0b_r}{g_{i,j}}\frac{TQ}{60}
\end{aligned}
\end{equation}
where $b_r$ and $C_r$ are the bandwidth of each RF channel and the achievable date rate allocated to real-time communication respectively.

\section{Problem formulation}

\subsection{Single UAV System}

In the case that all the ground users are served by one UAV base station, the probability that a channel is QoS is decided by the elevation angle (or in other words, the horizontal and vertical distance between the user and the UAV base station) and the ambient environment. There is obviously a trade-off between the free space path loss and the probability of LoS, so that we can find an optimal altitude for the UAV base station to save energy.

In the conventional cellular wireless communication system, the whole bandwidth is evenly distributed into multiple RF channels. In this project, we are trying to find a method to categorize the users into several groups based on the channel quality between each user and the UAV base station. The available bandwidth is distributed to each user group depending on the channel quality of users in each group.

The overall energy consumption in this system $E_{tot,1}$ which is the summation of propulsion energy and communication related energy can be written as \eqref{Etot1}.

\newcounter{mytempeqncnt}
\begin{figure*}[!t]
\normalsize
\setcounter{mytempeqncnt}{\value{equation}}
\vspace{-.5em}
\begin{equation}
\label{Etot1}
\begin{aligned}
E_{tot,1}&=E_{c}+E_{n}+E_{r}+E_{UAV}\\
&=\sum_{i=1}^n(2^{\frac{C_{c}}{b_c}}-1)\frac{N_0b_cLn\lambda}{\overline{g}_{i,1}C_{c}}+\sum_{\xi=1}^{k_n} \sum_{i\in \mathcal{R}_\xi}(2^{\frac{C_{\xi}}{b_\xi}}-1)\frac{N_0b_\xi}{\overline{g}_{i,1}}+\sum_{i=1}^n (2^{\frac{C_{r}}{b_{r}}}-1)\frac{N_0b_r}{\overline{g}_{i,1}}\frac{TQ}{60}\\
&+\frac{\delta}{8}\rho sA_u\Omega^3R_u^3+(1+k)\frac{W^{3/2}}{\sqrt{2\rho A_u}}
\end{aligned}
\end{equation}
\vspace{-.5em}
\begin{equation}
\label{Etot2}
\begin{aligned}
E_{tot,2}&=E_{c}+E_{n}+E_{r}+E_{UAV}\\
&=\sum_{j=1}^2 \sum_{i\in \mathcal{X}_j}(2^{\frac{C_{c}}{b_c}}-1)\frac{N_0b_cLn\lambda}{\overline{g}_{i,j}C_{c}}+
\sum_{\xi=1}^{k_n/2} \sum_{i\in \mathcal{R}_\xi}(2^{\frac{C_{\xi}}{b_\xi}}-1)\frac{N_0b_\xi}{\overline{g}_{i,1}}+\sum_{\xi=\frac{k_n}{2}+1}^{k_n} \sum_{i\in \mathcal{R}_\xi}(2^{\frac{C_{\xi}}{b_\xi}}-1)\frac{N_0b_\xi}{\overline{g}_{i,2}}\\
&+\sum_{j=1}^2 \sum_{i\in \mathcal{X}_j}(2^{\frac{C_{r}}{b_{r}}}-1)\frac{N_0b_r}{\overline{g}_{i,j}}\frac{TQ}{60}+E_{UAV}+\frac{\delta}{4}\rho sA_u\Omega^3R_u^3+(1+k)\frac{2W^{3/2}}{\sqrt{2\rho A_u}}
\end{aligned}
\end{equation}
\hrulefill
\vspace*{-4pt}
\vspace{-1em}
\end{figure*}

For the purpose of minimizing the energy consumed by this system, the optimization problem can be written mathematically as:

\begin{equation}
\begin{aligned}
\label{o2}
\mathop{\min}\limits_{b_c,b_\xi,b_r,C_c,C_\xi,C_r} &E_{tot,1}\\
\rm{Subject\:to :}&\:\: b_c+\sum_{\xi=1}^{k_n}b_\xi+k_r b_r \leq B,\\
&\:\: 0<b_c<B,\\
&\:\: 0<b_r<B,\\
&\:\: 0<b_\xi<B,\:\:\: \forall \xi \in \{1,2,\cdots,k_n\}\\
&\:\: C_{req}\leq C_c,\\
&\:\: R_{b}\leq C_r,\\
&\:\: R_n\leq C_\xi,\:\:\: \forall \xi \in \{1,2,\cdots,k_n\}\\
\end{aligned}
\end{equation}

It can be proved that this optimization problem is convex. The optimal solution can be obtained efficiently.

\begin{theorem}
Problem (\ref{o2}) is a convex problem.
\end{theorem}
\itshape {Proof:}  \upshape
Here, $b_c$ and $C_c$ are the only variables in $E_C$. Similarly, $b_\xi$ and $C_\xi$ are the only variables in $E_n$, $b_r$ and $C_r$ are the only variables in $E_r$. The second derivatives of $E_c$, $E_n$ and $E_r$ are all positive. That is, $E_c$, $E_n$ and $E_r$ are all convex. The summation of there three convex functions, $E_{tot,1}$, is still a convex function. Meanwhile, all the inequality constraints in Problem (\ref{o2}) are convex. Therefore, optimization problem (\ref{o2}) is a convex problem.

\subsection{Double UAV System}

By deploying multiple UAVs instead of one single UAV, the quality of the channels between the users and the UAV base station can be improved significantly. Therefore, the energy used for transmitting messages is reduced, which may cancel out the propulsion energy cost by additional UAVs.

The major problem is to find out the optimal matches between the ground users and the UAVs. The difference of all the users’ channel gains to both UAVs, i.e. $\overline{g}_{i,1}-\overline{g}_{i,2}$ are calculated and are sorted from largest to smallest. The first half users, i.e. $\mathcal{X}_1$ in this list are selected to be served by UAV $1$ while the second half, i.e. $\mathcal{X}_2$ are chosen to be connected to UAV $2$.

The users are also needed to be categorized into multiple groups for the purpose of transmitting non-real-time messages. If the number of user groups $k_n$ remains unchanged, then the first $k_n/2$ sets, i.e. $\mathcal{R}_1$ to $\mathcal{R}_{k_n/2}$ are all composed of the users connecting to UAV $1$ and the second half of user sets, i.e. $\mathcal{R}_{k_n/2+1}$ to $\mathcal{R}_{k_n}$ contains the ones that are served by UAV $2$. 

Simple to the single-UAV case, we can write the overall energy consumed by all the users $E_{tot,2}$ as \eqref{Etot2}.

In order to minimize the energy consumed by this system, we can write the optimization problem as follows:

\begin{equation}
\begin{aligned}
\label{o3}
\mathop{\min}\limits_{b_c,b_\xi,b_r,C_c,C_\xi,C_r}
&E_{tot,2}\\
\rm{Subject\:to :}&\:\: b_c+\sum_{\xi=1}^{k_n}b_\xi+k_r b_r \leq B,\\
&\:\: 0<b_c<B,\\
&\:\: 0<b_r<B,\\
&\:\: 0<b_\xi<B,\:\:\: \forall \xi \in \{1,2,\cdots,k_n\},\\
&\:\: C_{req}\leq C_c,\\
&\:\: R_{b}\leq C_r,\\
&\:\: R_n\leq C_\xi,\:\:\: \forall \xi \in \{1,2,\cdots,k_n\}\\
\end{aligned}
\end{equation}

Similar to Problem (\ref{o2}), Problem (\ref{o3}) can be proved to be a convex problem as well. Hence, we can obtain the optimal solution of this problem efficiently by using the conventional methods such as dual method in \cite{8379427,8764580}.

\vspace{-.25em}
\section{Numerical Results}\vspace{-.5em}

For our simulations, 
the side length of the region where the ground users are distributed is set as $R=2000$m and the UAVs are assumed to be in the center of their coverage. The average transmission rate, packet length and required latency of control messages are set as $\lambda=10$, $L=200$bits, and $T_{req}=1$ms respectively. The average data rate of non-real-time service is $R=1$Mbps for each users. The average call duration and average number of calls per hour of real-time communication are set as $T=2$minutes, and $Q=10$ respectively. We set maximum blocking probability $B_l=2\%$. The burst data rate $R_b$ is set identical to the burst data rate of LTE network, which equals to 270.8Kbps. The AWGN spectral density and carrier frequency are set as $N_0=10^{-16}$W/Hz and $f_c=1$GHz. The UAV's propulsion energy related parameters are set as what is showed in table \ref{propulsion energy}.

\begin{figure*}[!t]
\centering
\subfigure[Energy save rate in percent]{
\begin{minipage}[t]{0.45\linewidth}
\centering
\includegraphics[width=\linewidth]{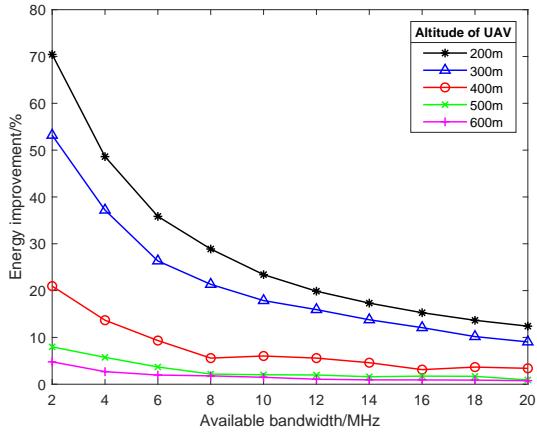}
\end{minipage}
}
\subfigure[Energy consumption comparison]{
\begin{minipage}[t]{0.45\linewidth}
\centering
\includegraphics[width=\linewidth]{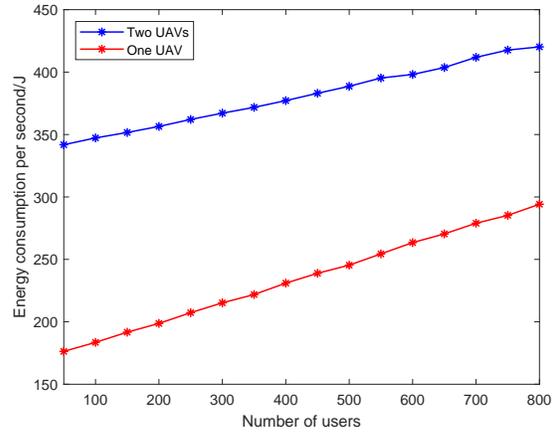}
\end{minipage}
}

\centering
\caption{Energy consumption in suburban region}
\label{fig1}
\end{figure*}

\begin{table}[H]
\centering
\caption{UAV's propulsion energy related parameters}
\label{propulsion energy}
\begin{tabular}{ c|c|c } 
 \hline
Physical meaning&Notation&Simulation value\\ 
 \hline\hline
Aircraft weight in Newton&$W$&20\\ 
\hline
Air density in kg/m$^3$&$\rho$&1.225\\
\hline
Rotor radius in meter&$R_u$&0.4\\
\hline
Rotor disc area in m$^2$&$A_u$&0.503\\
\hline
Blade angular velocity in radians/s&$\Omega$&300\\
\hline
Rotor solidity&$s$&0.05\\
\hline
Incremental correction factor&$k$&0.1\\
\hline
Profile drag coefficient&$\delta$&0.012\\
\hline
\end{tabular}
\end{table}

\begin{figure*}[!t]
\centering
\subfigure[Energy save rate in percent]{
\begin{minipage}[t]{0.45\linewidth}
\centering
\includegraphics[width=\linewidth]{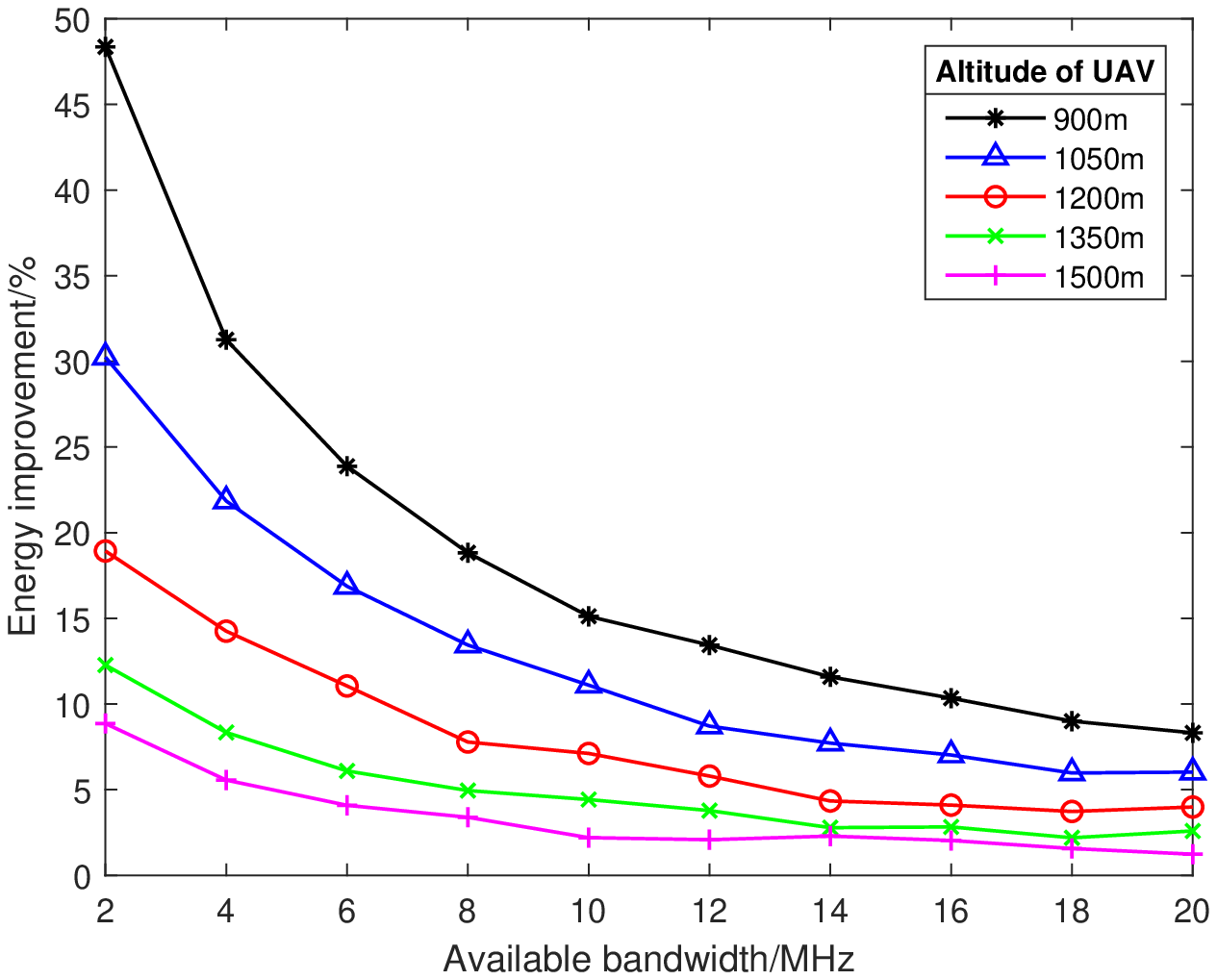}
\end{minipage}
}
\subfigure[Energy consumption comparison]{
\begin{minipage}[t]{0.45\linewidth}
\centering
\includegraphics[width=\linewidth]{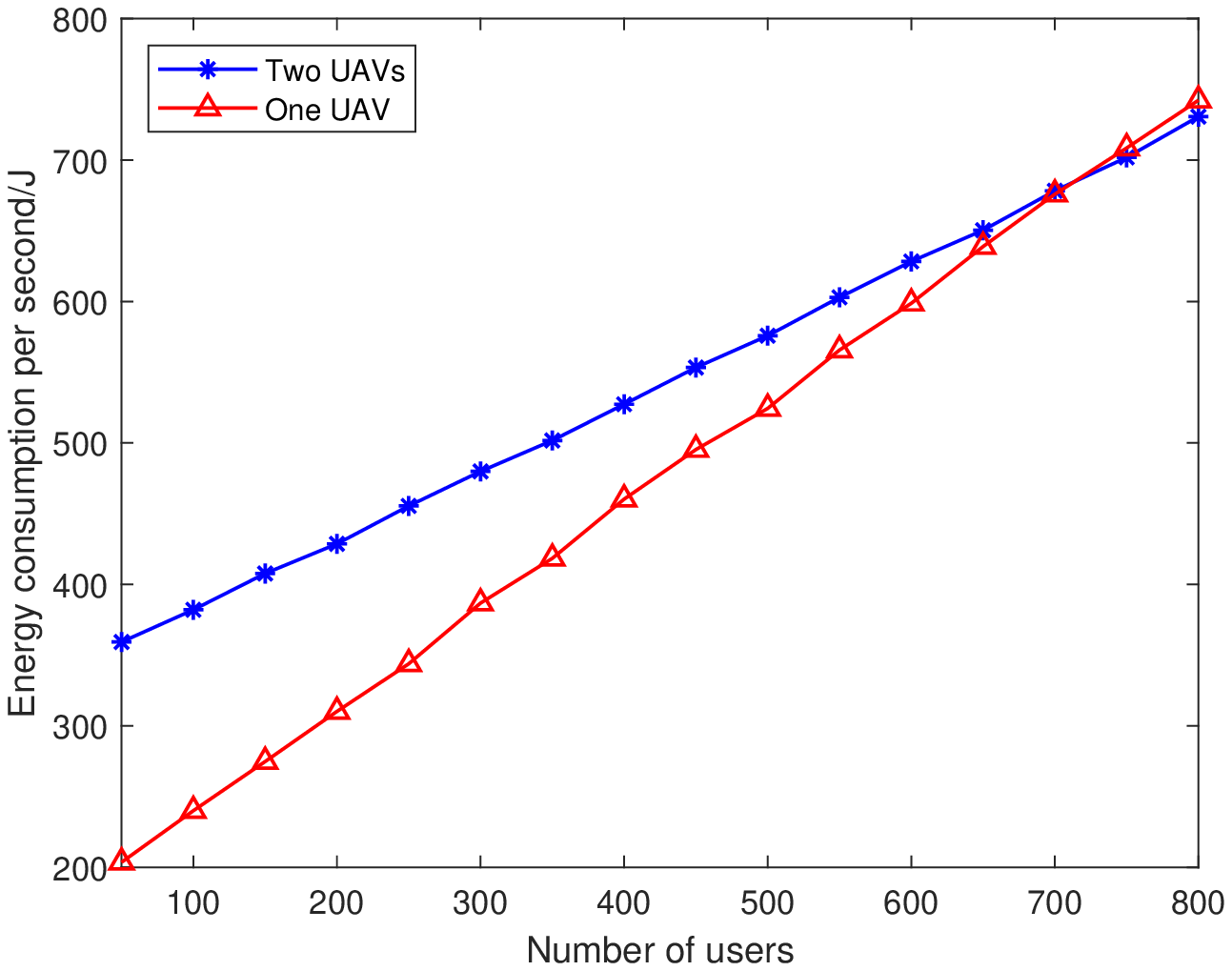}
\end{minipage}
}
\centering
\caption{Energy consumption in urban region}
\label{fig2}
\end{figure*}

\begin{figure*}[!t]
\centering
\subfigure[Energy save rate in percent]{
\begin{minipage}[t]{0.45\linewidth}
\centering
\includegraphics[width=\linewidth]{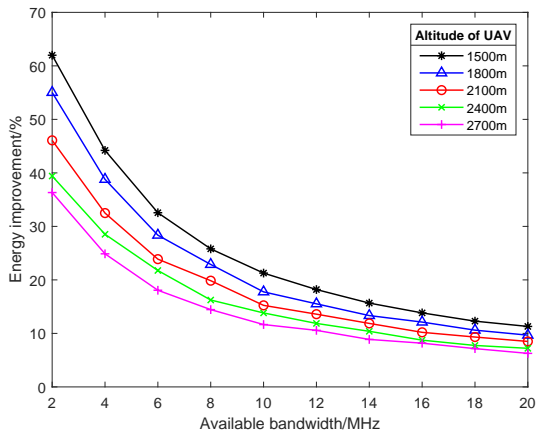}
\end{minipage}
}
\subfigure[Energy consumption comparison]{
\begin{minipage}[t]{0.45\linewidth}
\centering
\includegraphics[width=\linewidth]{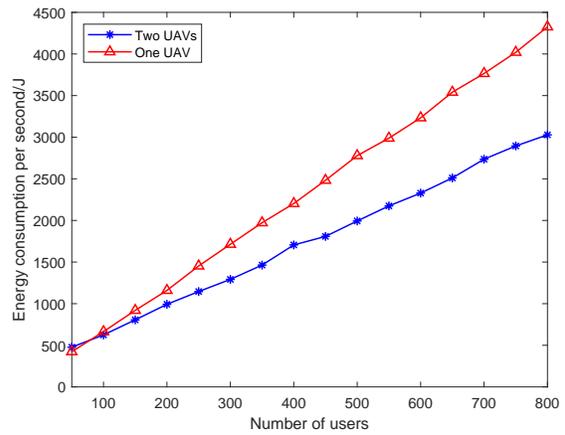}
\end{minipage}
}
\centering
\caption{Energy consumption in dense urban region}
\label{fig3}
\end{figure*}

Figs. \ref{fig1}-\ref{fig3} show the energy efficiency which can be calculated by the percentage of energy saved by grouping users and the comparison between the system served by one and two UAVs in three kinds of environments respectively.

We can observe from the graphs that more energy can be saved if the altitude of the UAV is lower. This happens because the probability of LoS reduces as the altitude of UAV reduces. The decreasing of LoS probability will further leads to the larger differences between the qualities of the channels.

In suburban region, the quality of the channels is rather high so that the communication related energy is much less than the propulsion energy consumed in this system. Therefore, it is better to deploy only one UAV base station for the purpose of saving energy.

It is remarkable that only when more than 700 users are served, deploying two UAVs simultaneously can be more energy-efficient in urban region.

In dense urban regions, more energy is consumed by the users to transmit messages.  As long as more than 100 users are in the system, more than one UAVs should be deployed in order to save energy.

\vspace{-.25em}
\section{Conclusion}
\vspace{-.5em}
In this paper, we aim to allocate the overall bandwidth to multiple RF channels to minimize the total energy consumption of this system while satisfying all three services of the system. If the bandwidth resource is limited, there are ways to enhance the energy efficiency of the system. The first method is to categorize the users into multiple groups and offer each user group a unique RF channel with different bandwidth. Numerical results show that plenty of energy can be saved through using this technique. Moreover, deploying multiple UAVs can sometime result in higher energy efficiency compared with deploying one UAV only. It is proved through the comparison between the same user sets served by one UAV and two UAVs.

\section*{Acknowledgment}
This work was supported by the Engineering and
Physical Science Research Council (EPSRC) through the
Scalable Full Duplex Dense Wireless Networks (SENSE)
grant EP/P003486/1.


\vspace{-0.5em}
\bibliographystyle{IEEEtran}
\bibliography{IEEEabrv,MMM}

\end{document}